\documentclass[aps,prl,twocolumn,unsortedaddress]{revtex4-1}
\usepackage{epsfig,pstricks,graphicx}
\usepackage{amsmath,amssymb}
\usepackage[mathscr]{eucal}
\usepackage{color}
\usepackage[normalem]{ulem}

\usepackage{dsfont}
\newcommand{\bra}[1]{\ensuremath{\langle #1 |}}
\newcommand{\ket}[1]{\ensuremath{| #1 \rangle}}
\newcommand{\tr}{\operatorname{tr}}
\def\id{{\rm 1\kern-.22em l}}
\def\CC{{\rm\kern.24em \vrule width.04em height1.46ex depth-.07ex
         \kern-.30em C}}
\def\RR{{\rm
         \vrule width.04em height1.58ex depth-.0ex
         \kern-.04em R}}

\newcommand{\rme}{\ensuremath{\mathrm{e}}}
\newcommand{\rmi}{\ensuremath{\mathrm{i}}}

\begin{document}

\title{Negativity as a counter of entangled dimensions}
\author{Christopher Eltschka}
\affiliation{Institut f\"ur Theoretische Physik, 
         Universit\"at Regensburg, D-93040 Regensburg, Germany}
\author{Jens Siewert}
\affiliation{Departamento de Qu\'{\i}mica F\'{\i}sica, Universidad del Pa\'{\i}s Vasco UPV/EHU,
             E-48080 Bilbao, Spain}
\affiliation{IKERBASQUE, Basque Foundation for Science, E-48011 Bilbao, Spain}
\begin{abstract}
Among all entanglement measures negativity
arguably is the best known and most popular tool
to quantify bipartite quantum correlations.
It is easily computed for arbitrary states
of a composite system and can therefore be applied to discuss
entanglement in an ample variety of situations. However, 
its direct physical meaning has not been pointed out yet.
We show that the negativity can be viewed as an estimator of
how many degrees of freedom
of two subsystems are entangled. As it is possible to give lower bounds
for the negativity even in a device-independent 
setting, it is the appropriate quantity to certify quantumness of both 
parties in a bipartite system and to determine the minimum number
of dimensions that contribute to the quantum correlations.
\end{abstract}
\maketitle

%
%
%%%%%%%%%%%%%%%%%%%%%%%%%%%%%%%%%%%%%%%%%%%%%%%%%%%%%%%%%%%%%%%%%%%%%%%%%%%%%%%%%%%%%%%%%%
{\em Introduction. -- } 
%%%%%%%%%%%%%%%%%%%%%%%%%%%%%%%%%%%%%%%%%%%%%%%%%%%%%%%%%%%%%%%%%%%%%%%%%%%%%%%%%%%%%%%%%%
The dimension, that is, the number of independent degrees of freedom 
is a particularly important system parameter.
It is relevant, for example, for
the security of cryptography schemes and for the significance of 
Bell inequality violation~\cite{Acin2006,Brunner2008}. In general,
in information processing (both classical and quantum) the dimensionality
may be regarded as a resource and is therefore crucial for system
performance.

The device-independent characterization of physical
systems~\cite{Acin2006,Brunner2008,Gallego2010,Bancal2011,Pal2011,Bancal2012,Acin2012,Cabello2012,Moroder2013} 
without a priori restrictions regarding the underlying
structure of mathematical models is fundamental for our understanding of Nature.
The goal %of device-independent characterization 
is to obtain the desired physical information based only on the statistics 
from certain measurement outcomes ('prepare and measurement scenario', 
Ref.~\cite{Gallego2010})
without reference to internal properties or mechanisms of a device.
In recent years numerous schemes for device-independent dimension
testing and other system properties have been proposed. 
There are methods that
detect the minimum number of classical or quantum degrees of
freedom for a single system~\cite{Gallego2010,Acin2012,Cabello2012}.
The dimensionality may be inferred also from Bell-inequality
violation~\cite{Acin2006,Brunner2008}. On the other hand, there
are device-independent methods for multipartite 
entanglement detection~\cite{Bancal2011,Pal2011,Bancal2012,Moroder2013}.
In our work we propose direct counting of entangled dimensions based on a well-known
entanglement measure for bipartite systems, the negativity, thereby
elucidating the physical meaning of the latter. The method can be made
device independent by invoking techniques from Refs.~\cite{Bancal2012,Moroder2013}.
With our result we cannot draw any conclusion regarding the classical dimensions 
of the two local systems. However, since entanglement is possible only between quantum
degrees of freedom we directly obtain the minimum number of quantum levels for
both parties which then are certified to be quantum without further assumption.

To demonstrate this we first study a nontrivial family of mixed states that
can be defined for any $d\times d$-dimensional bipartite system, the
{\em axisymmetric states}. Their negativity provides a clear illustration
for the central statement of our article. It is then easy to show that
this statement holds for arbitrary states as well. Finally we establish
the link to the device-independent description that concludes our
construction of a device-independent bound on the number of entangled
dimensions for two-party systems.

%%%%%%%%%%%%%%%%%%%%%%%%%%%%%%%%%%%%%%%%%%%%%%%%%%%%%%%%%%%%%%%%%%%%%%%%%%%%%%%%%%%%%%%%%%
{\em Negativity. -- } 
%%%%%%%%%%%%%%%%%%%%%%%%%%%%%%%%%%%%%%%%%%%%%%%%%%%%%%%%%%%%%%%%%%%%%%%%%%%%%%%%%%%%%%%%%%
The negativity was first 
used by Zyczkowski {\em et al.}~\cite{Zyczkowski1998} and 
subsequently introduced as an entanglement measure
by Vidal and Werner~\cite{VidalWerner2002}.
Consider the state $\rho$ of a bipartite system with finite-dimensional
Hilbert space $\mathcal{H}_A\otimes\mathcal{H}_B$. The negativity
is defined as
\begin{equation}
   \mathcal{N}(\rho)\ =\ \frac{1}{2}\left( ||\rho^{T_A}||_1-1 \right)
\end{equation}
where $\rho^{T_A}$ denotes the partial transpose with respect to party $A$
and $||M||_1\equiv\tr\sqrt{M^{\dagger}M}$ is the trace norm of the matrix $M$.
We slightly modify this definition by
introducing the quantity
\begin{equation}
   \mathcal{N}_{\text{dim}}(\rho)\ =\ 2\ \mathcal{N(\rho)}+1\ \equiv\ ||\rho^{T_A}||_1\ \ .
\end{equation}
As our discussion proceeds it will turn out that the least integer
greater than or equal to
$\mathcal{N}_{\text{dim}}$ is a lower bound to the number of entangled dimensions 
between the parties $A$ and $B$.

%%%%%%%%%%%%%%%%%%%%%%%%%%%%%%%%%%%%%%%%%%%%%%%%%%%%%%%%%%%%%%%%%%%%%%%%%%%%%%%%%%%%%%%%%%
{\em Axisymmetric states. -- } 
%%%%%%%%%%%%%%%%%%%%%%%%%%%%%%%%%%%%%%%%%%%%%%%%%%%%%%%%%%%%%%%%%%%%%%%%%%%%%%%%%%%%%%%%%%
In studies of entanglement properties
it is often useful to define
families of states with a certain symmetry~\cite{Vollbrecht2001}, such as 
the Werner states~\cite{Werner1989} and the isotropic states~\cite{Horodecki1999}.
Here we introduce the {\em axisymmetric states} $\rho^{\text{axi}}$
for two qudits
which are all the states obeying the same symmetries as the 
maximally entangled state in $d$ dimensions
\begin{equation}
   \ket{\Psi_d}\ =\ \frac{1}{\sqrt{d}}\left(\ket{11}+\ket{22}+\ldots +\ket{dd}
                                    \right)\ \ ,
\label{eq:Bell}
\end{equation}
that is
\begin{itemize}
\item exchange of the two qudits,
\item simultaneous permutations of the basis states for both qudits
 {\em e.g.}, $\ket{0}_A\leftrightarrow\ket{1}_A$ and $\ket{0}_B\leftrightarrow\ket{1}_B$,
\item coupled phase rotations
  \begin{equation}
    \label{eq:zrot}
    U(\varphi_1,\ldots,\varphi_{d-1})=
   \rme^{\rmi\sum_j\varphi_j\mathfrak{g}_j}
   \otimes\rme^{-\rmi\sum_k\varphi_k\mathfrak{g}_k}
   \nonumber
  \end{equation}
where $\mathfrak{g}_j$ ($j=1,\ldots,d-1$) are the diagonal 
generators of the group SU($d$).
\end{itemize}
Apart from the maximally entangled state Eq.~\eqref{eq:Bell} this family contains 
only (mostly full-rank) mixed states. For any $d\geqq 2$  these states
are given by two real parameters $x$ and $y$ that describe the position of the
state in a plane triangle (in close analogy to the Greenberger-Horne-Zeilinger
symmetric states~\cite{Eltschka2012}), see Fig.~1. In order to determine
the lengths of the triangle sides we choose the Euclidean metric of the
triangle to
coincide with the Hilbert-Schmidt metric of the density matrices.
This enables us to deduce various physical facts from Fig.~1 
merely by means of geometric intuition.

Axisymmetric states for $d\times d$ systems can be represented as $d^2\times d^2$
matrices with diagonal elements
\[
          \rho^{\text{axi}}_{jj,jj} \ =\ \frac{1}{d^2} + a \  ,\ \ \ 
          \rho^{\text{axi}}_{jk,jk} \ =\ \frac{1}{d^2}-\frac{a}{d-1}\ \ (j\neq k)
\]
($j,k=1,\ldots,d$) and off-diagonal entries
\[
          \rho^{\text{axi}}_{jj,kk} \ =\ b\ \ (j\neq k) \ \ ,
\]
all other off-diagonal elements vanish. 
The ranges for the matrix elements are
\begin{align}\label{eq:range-a}
   -\frac{1}{d^2}\ & \leqq\ a\ \leqq\ \frac{d-1}{d^2}\\
\label{eq:range-b}
   -\frac{1}{d-1}\left(\frac{1}{d^2}+a\right)\ & \leqq\ b\ \leqq\ \left(\frac{1}{d^2}+a\right) \ \ .
\end{align}
From Eqs.~\eqref{eq:range-a}, \eqref{eq:range-x} we recognize the triangular shape
of the set of axisymmetric states. With this choice of parametrization the 
fully mixed state $\frac{1}{d^2}\id_{d^2}$ is located at the origin.

Now we choose the scale of $a\equiv \alpha y$ and $b\equiv \beta x$ such that the Euclidean metric
for $x$ and $y$ with the Hilbert-Schmidt metric in the space
of density matrices. We define the Hilbert-Schmidt scalar product of two
matrices $M_1$ and $M_2$ as
$\langle M_1, M_2 \rangle_{\text{HS}}
  \equiv
  \tr\left(M_1^{\dagger}M_2\right)$.
With this we find $\alpha=\frac{\sqrt{d-1}}{d}$ and $\beta=\sqrt{d(d-1)}^{-1}$ so that
\begin{align}\label{eq:range-y}
   -\frac{1}{d\sqrt{d-1}}\ & \leqq\ y\ \leqq\ \frac{\sqrt{d-1}}{d}\\
\label{eq:range-x}
   -\frac{1}{\sqrt{d(d-1)}} \ & \leqq\ x\ \leqq\ 
    \sqrt{\frac{d-1}{d}}    \ \ .
\end{align}
%

%
%%%%%%%%%%%%%%%%%%%%%%%%%%%%%%%%%%%%%%%%%%%%%%%%%%%%%%%%%%%%%%%%%%%%%
%% FIGURE 1
%%%%%%%%%%%%%%%%%%%%%%%%%%%%%%%%%%%%%%%%%%%%%%%%%%%%%%%%%%%%%%%%%%%%%
\begin{figure}[htb]
  \centering
  \includegraphics[width=.97\linewidth]{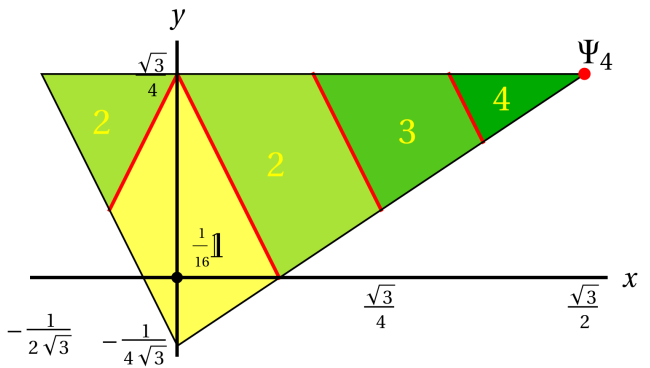}
  \caption{The convex set of $d\times d$ axisymmetric states
           $\rho^{\text{axi}}$, here for $d=4$.
              The family is characterized
              by two real parameters. While 
              $x$ is proportional to the off-diagonal element, 
              $y$ describes the asymmetry between the two types of
              diagonal elements (see Section Methods).
              The upper right corner corresponds to the
              maximally entangled state $\ket{\Psi_d}$ (the only
              pure state), the completely mixed state 
              $\frac{1}{d^2}\id_{d^2}$ resides at the origin.\\
              The states with local dimension $d$ have $d$ SLOCC 
              classes corresponding to their Schmidt number $k$
              (indicated by the yellow numbers in the regions). 
              The states with Schmidt number $\leqq k$
              form the convex sets $S_k$ and build a hierarchy
              $S_1\subset S_2\subset\ldots\subset S_d$.
              Note that Schmidt number $k=1$ corresponds
              to separable states which are considered classical.
    }
  \label{fig:2Ddreieck}
\end{figure}
%%%%%%%%%%%%%%%%%%%%%%%%%%%%%%%%%%%%%%%%%%%%%%%%%%%%%%%%%%%%%%%%%%%%%
%

%%%%%%%%%%%%%%%%%%%%%%%%%%%%%%%%%%%%%%%%%%%%%%%%%%%%%%%%%%%%%%%%%%%%%%%%%%%%%%%%%%%%%%%%%%
{\em Entanglement of axisymmetric states. -- } 
%%%%%%%%%%%%%%%%%%%%%%%%%%%%%%%%%%%%%%%%%%%%%%%%%%%%%%%%%%%%%%%%%%%%%%%%%%%%%%%%%%%%%%%%%%
Remarkably, many entanglement properties of axisymmetric states can be 
determined
exactly. The entanglement class of a bipartite state with respect to 
stochastic local operations and classical communication (SLOCC) 
is given by its Schmidt number, the minimal required Schmidt rank 
for any pure-state decomposition of the state. By using the optimal 
Schmidt number witnesses~\cite{Lewenstein2001}
\begin{equation}
\mathcal{W}\ =\ \frac{k-1}{d}\ \id_{d^2}-\ket{\Psi_d}\!\bra{\Psi_d}
\nonumber
\end{equation}
($2\leqq k\leqq d)$
we find for each state $\rho^{\text{axi}}(x,y)$ the corresponding
Schmidt number, cf.~Fig.~1. Notably, the borders between the 
SLOCC classes for $x\geqq 0$ are %given by 
straight lines % that are 
parallel 
to the lower left side of the triangle. This is no surprise since those
lines correspond to states of constant overlap with the maximally entangled
state $\ket{\Psi_d}$. Moreover, we easily identify the compact convex 
sets $S_k$ of states with Schmidt number {\em at most} 
equal to $k$~\cite{Lewenstein2001}.

In the next step, we calculate  the negativity for axi\-symmetric states. 
To this end we consider the
eigenvalue problem for the partial transpose of $\rho^{\text{axi}}$. It results
in $\frac{d(d-1)}{2}$ identical eigenvalue problems for $2\times 2$ matrices
\[     \begin{pmatrix} \frac{1}{d^2}-\frac{a}{d-1} & b \\
                       b & \frac{1}{d^2}-\frac{a}{d-1} 
       \end{pmatrix}
\]
which have the eigenvalues 
\[
    \lambda_{\pm}\ =\ \frac{1}{d^2}-\frac{a}{d-1} \ \pm \ |b|\ \ .
\]
Adding the absolute negative eigenvalues and rewriting $a$ and $b$ in terms
of $x$ and $y$ leads to
\begin{equation}
    \mathcal{N} = \max\left\{ 0 , \frac{1}{2}\left( \sqrt{d(d-1)}|x|+\sqrt{d-1}y-\frac{d-1}{d}
                                             \right)
                       \right\}\ .
\end{equation}
From this we find the exact $\mathcal{N}_{\text{dim}}$
for the entangled axisymmetric states 
\begin{equation}
\mathcal{N}_{\text{dim}}(\rho^{\text{axi}}(x,y))\ =\ \sqrt{d(d-1)}|x|+\sqrt{d-1}y
                                           +\frac{1}{d}
\label{eq:axi-negativity}
\end{equation}
which is noteworthy in several respects. First, the negativity is
a {\em linear} function of $|x|$ and $y$ (see Fig.~2).
%%so that the graph of the
%%function is represented by planes, see Fig.~2. 
A state has nonvanishing
negativity if and only if it is not separable. Consequently, 
there are no entangled axisymmetric states with positive partial transpose.
Further, and most importantly, the borders between SLOCC classes 
correspond to isolines for integer values 
of the negativity. With the ceiling function $\lceil x\rceil$, 
the smallest integer greater than or equal to $x$, we see 
that for axisymmetric states $\rho^{\text{axi}}(x,y)$
\begin{equation}
   \text{SLOCC\ class}\ k\  
            \ =\ 
            \lceil\mathcal{N}_{\text{dim}}(x,y)\rceil\ \ .
\end{equation}
However, the SLOCC class, that is, the minimum required
Schmidt rank of the 
pure states in the decomposition of $\rho^{\text{axi}}$ 
counts the number of degrees of freedom in which subsystems
$A$ and $B$ are entangled. In consequence our result implies
that for axisymmetric states
the modified negativity $\mathcal{N}_{\text{dim}}$ is a precise counter
of entangled dimensions. 

%%%%%%%%%%%%%%%%%%%%%%%%%%%%%%%%%%%%%%%%%%%%%%%%%%%%%%%%%%%%%%%%%%%%%%%%%%%%%%%%%%%%%%%%%%
{\em Dimension estimator for arbitrary states. -- } 
%%%%%%%%%%%%%%%%%%%%%%%%%%%%%%%%%%%%%%%%%%%%%%%%%%%%%%%%%%%%%%%%%%%%%%%%%%%%%%%%%%%%%%%%%%
Naturally the question is imposed to which
extend this statement holds for all bipartite states.
Due to the existence of entangled states with positive 
partial transpose~\cite{Horodecki1998} it is 
clear that the negativity cannot be a precise counter of
entangled dimensions for arbitrary states.
In the following we prove that, while not being an exact counter,
the modified negativity $\mathcal{N}_{\text{dim}}$ is always a lower bound to 
the Schmidt number.

To this end, we explicitly show again how to calculate the negativity 
for pure entangled states of Schmidt rank $k$. 
Any such state is locally equivalent to 
$\ket{\Psi_k}$, the maximally entangled state of Schmidt rank $k$.
Considering the partial transpose of $\ket{\Psi_k}\!\bra{\Psi_k}$
\begin{equation}
\ket{\Psi_k}\!\bra{\Psi_k}=\frac{1}{k}\sum_{\alpha \beta}
                                                 \ket{\alpha \alpha }\!\bra{\beta\beta}
\ \ \stackrel{T_A}{\longrightarrow}\ \ 
                   \frac{1}{k}\sum_{\alpha \beta}\ket{\beta\alpha }\!\bra{\alpha \beta}
\nonumber
\end{equation}
it is evident that $\mathcal{N}_{\text{dim}}(\Psi_k)=2\frac{1}{k}\frac{k(k-1)}{2}+1=k$.
Now, since according to Ref.~\cite{VidalWerner2002}
the negativity is a convex function of the 
state $\rho$ we find for an arbitrary state of Schmidt number $k$
\begin{equation}
\mathcal{N}_{\text{dim}}(\rho)\ \leqq\  \sum_j p_j\mathcal{N}_{\text{dim}}(\psi_j)\ \leqq 
                         \ \sum_j p_j k\  =\  k
\label{eq:estimate}
\end{equation}
for $\rho=\sum_j p_j\ket{\psi_j}\!\bra{\psi_j}$
as in that case $\mathcal{N}_{\text{dim}}(\psi_j)\leqq k$.
We mention that these estimates are valid for
arbitrary bipartite systems  with $d\times d^{\prime}$  dimensions,
both for $d=d^{\prime}$ and for $d\neq d^{\prime}$. This is because
the Schmidt rank of a pure $d\times d^{\prime}$ state cannot exceed
the smaller of the two local dimensions.
This concludes the proof that the modified negativity
$\mathcal{N}_{\text{dim}}$ is an estimator for the number of entangled
dimensions of arbitrary two-party states. $\Box$
%
%%%%%%%%%%%%%%%%%%%%%%%%%%%%%%%%%%%%%%%%%%%%%%%%%%%%%%%%%%%%%%%%%%%%%
%% FIGURE 2
%%%%%%%%%%%%%%%%%%%%%%%%%%%%%%%%%%%%%%%%%%%%%%%%%%%%%%%%%%%%%%%%%%%%%
\begin{figure}[thb]
  \centering
  \includegraphics[width=.97\linewidth]{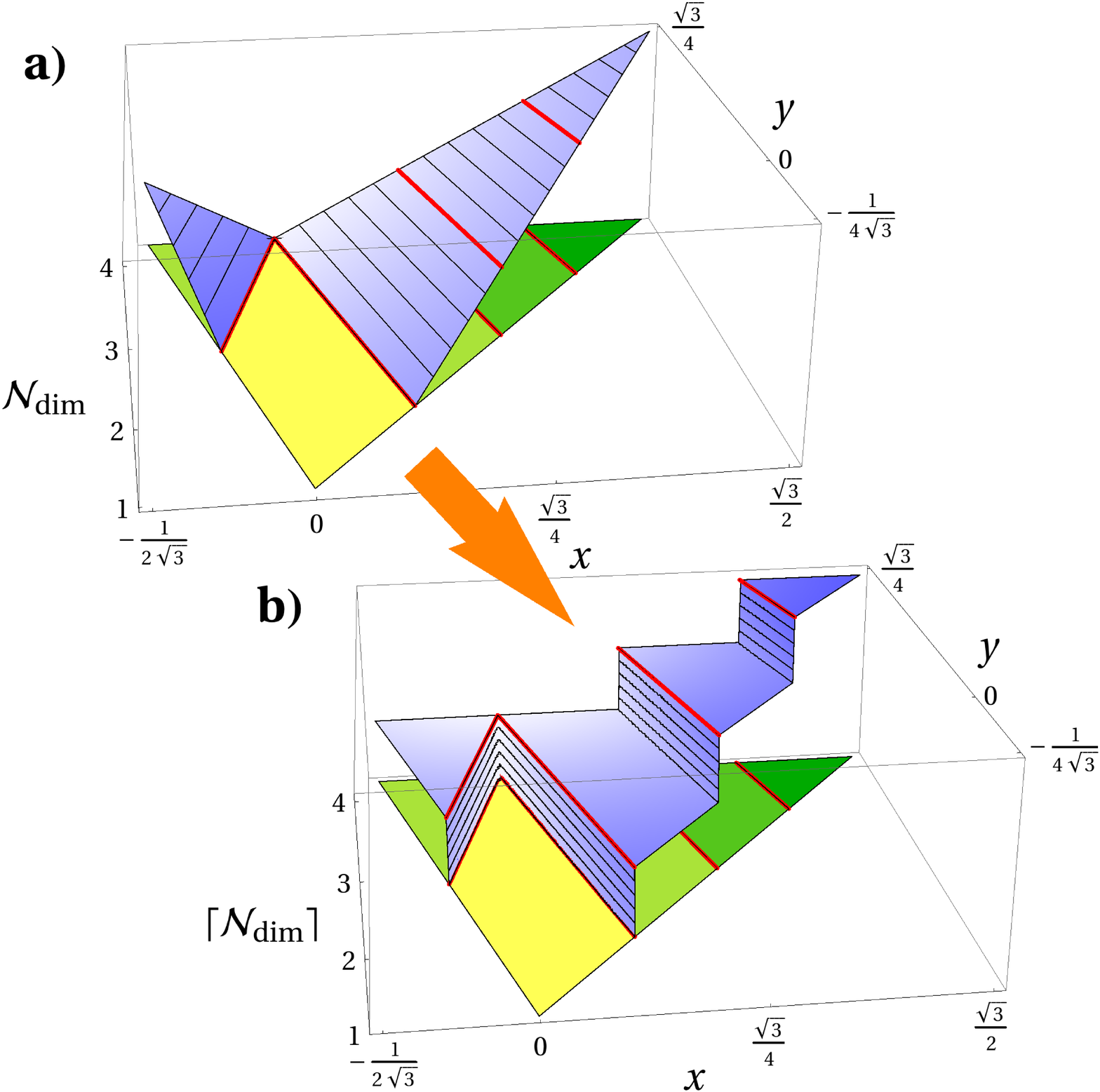}
  \caption{Exact modified negativity $\mathcal{N}_{\text{dim}}$
             for $d\times d$ axisymmetric states
             $\rho^{\text{axi}}$, again for $d=4$.\\
           a) The blue surface displays $\mathcal{N}_{\text{dim}}(x,y)$.
              It depends linearly on $|x|$ and $y$. Note that
              the borders between SLOCC classes (red lines
              in the $x,y$ plane) are projections of 
              integer-value isolines of the modified negativity.\\
           b) The ceiling function
              $\lceil\mathcal{N}_{\text{dim}}(x,y)\rceil$
              (blue surface)
              counts the Schmidt number of $\rho^{\text{axi}}(x,y)$.
    }
  \label{fig:3Dplot}
\end{figure}
%%%%%%%%%%%%%%%%%%%%%%%%%%%%%%%%%%%%%%%%%%%%%%%%%%%%%%%%%%%%%%%%%%%%%
%

%%%%%%%%%%%%%%%%%%%%%%%%%%%%%%%%%%%%%%%%%%%%%%%%%%%%%%%%%%%%%%%%%%%%%%%%%%%%%%%%%%%%%%%%%%
{\em Device-independent dimension estimate. -- }
%%%%%%%%%%%%%%%%%%%%%%%%%%%%%%%%%%%%%%%%%%%%%%%%%%%%%%%%%%%%%%%%%%%%%%%%%%%%%%%%%%%%%%%%%%
It remains to discuss that a lower bound on the entangled
dimensions via the negativity, or $\mathcal{N}_{\text{dim}}$, 
can be obtained in a device-independent setting.
This technique has rencently been worked out by 
Moroder {\em et al.}~\cite{Moroder2013} and we sketch only
the main idea here. A device-independent scenario implies
that a number of generalized measurements are carried out
on the subsystems $A$ and $B$.  While the detailed actions 
$A_i$, $B_j$
of the measurement devices on the true state $\rho_{AB}$
are unknown to the observers, the outcomes
for each party labeled by $i$ and $j$, are mutually exclusive. 
One also defines $A_0= \id_A$ and $B_0=\id_B$.  The observers 'see'
$\rho_{AB}$ only via their preparation-measurement setup,
and (partially) determine the % so-called 
Hermitian matrix            %of moments
\begin{equation}
\chi_{ij,kl}[\rho_{AB}] = \tr\left[\rho_{AB}\left(A_k^\dagger A_i
                                           \otimes B_l^\dagger B_j\right)\right]
\end{equation}
with orthonormal bases $\{\ket{i}_{\tilde{A}}\}$, $\{\ket{j}_{\tilde{B}}\}$ 
in the outcome spaces $\tilde{A}$ and $\tilde{B}$.
This matrix depends linearly on $\rho_{AB}$ and is positive
whenever the true state $\rho_{AB}$ is positive.
Correspondingly, whenever $\rho_{AB}^{T_A}$ is  positive,
 $\chi^{T_A}$ is  positive, too.

The possibility to estimate the negativity relies on 
its varia\-tio\-nal expression~\cite{VidalWerner2002}: 
$\mathcal{N}(\rho_{AB})=\min \{\tr \sigma:
\mbox{$\sigma^{T_A}\geqq0$},(\rho_{AB}-\sigma)^{T_A}\geqq0\}$.
The properties of $\chi$ mentioned above mean
that the conditions for minimization hold also for $\chi[\rho_{AB}]$ and
$\chi[\sigma]$. Moreover, the optimized quantity $\tr \sigma$ equals
$\chi_{0000}[\sigma]$. Therefore, minimising $\chi_{0000}[\sigma]$
over \emph{all} matrices 
$\chi$ consistent with the measurement outcomes(and the condition $\tr
\rho_{AB}=1$) will give a lower bound for the negativity $\mathcal{N}(\rho_{AB})$.

Evidently, our findings are useful to characterize
a test system with unknown quantum dimension. By entangling
it with an auxiliary system of known dimension and measuring
the negativity a lower bound to the number of quantum levels
in the test system can be found.

We conclude by mentioning that the results regarding the 
negativity hold also for the convex-roof extended negativity~\cite{Lee2003}
because it is the largest convex function that coincides
with the negativity on pure states~\cite{Uhlmann2010}.
However, while improving the estimate in Eq.~\eqref{eq:estimate}
the negativity would forfeit its most important asset, namely that it can be 
calculated easily.
      
%%%%%%%%%%%%%%%%%%%%%%%%%%%%%%%%%%%%%%%%%%%%%%%%%%%%%%%%%%%%%%%%%%%%%%%%%%%%%%%%%%%%%%%%%%
%\section*{Acknowledgements}

{\em Acknowledgements -- .}
%%%%%%%%%%%%%%%%%%%%%%%%%%%%%%%%%%%%%%%%%%%%%%%%%%%%%%%%%%%%%%%%%%%%%%%%%%%%%%%%%%%%%%%%%%
This work was funded by the German Research Foundation within 
SPP 1386 (C.E.), and by Basque Government grant IT-472 (J.S.).
The authors thank O.\ G\"uhne and Z.\ Zimboras for helpful 
remarks and J.\ Fabian and K.\ Richter for their support.
%

%%%%%%%%%%%%%%%%%%%%%%%%%%%%%%%%%%%%%%%%%%%%%%%%%%%%%%%%%%%%%%%%%%%%%%%%%%%%%%%%%%%%%%%%%%

%
%
%
\end{document}